\documentclass[11pt,twocolumn,twoside]{article}
\usepackage{fully3d}
\usepackage{multicol}
\usepackage{subfig}
\usepackage{float}

\begin{document}

\title{Image reconstruction from tissue scattered events for ${\beta}^{+}\gamma$ coincidences in Compton-PET} 

\author[1]{Satyajit Ghosh}
\author[1]{Pragya Das}

\affil[1]{Department of Physics,
          Indian Institute of Technology Bombay, Mumbai, INDIA}


\maketitle
\thispagestyle{fancy}


\begin{customabstract}
For long time non-pure beta emitters are avoided from PET imaging due to extra dose and increase in background from Compton scattering. But advent of high-resolution Compton camera system opens up new domain of imaging. Various non-pure beta emitters are formed as beam irradiation byproduct in therapy which can be used in online beam range verification. In this case, the number of usable counts for imaging is generally 1-3 order lesser than normal PET scan. On the other hand, we know that in human PET scanner, 30-60\% can be tissue scattered coincidences in 3D case containing 80\% single scattered events. In this work, we have investigated feasibility of imaging using only single scattered coincidences for non-pure beta emitters in a Compton-PET system. The locus of tissue scatter point can be reduced to in generally two points after using Compton cone from both ends of 511 keV detections. Finally, annihilation point is estimated using Compton cone of 1157 keV gamma and time-of-flight information for the 511 keV. We believe independent assessment of underlying activity from single scattered data sets will increase confidence in image interpretation. 
\end{customabstract}


\section{Introduction}
For long time non-pure beta emitter radioisotopes (e.g., $^{44m}$Sc,$^{94}$Tc,$^{14}$O,$^{68}$Ga,$^{124}$I,$^{10}$C) are not used in PET imaging. This is because of extra dose and Compton scattering background that the quasi-simultaneously emitted extra gamma ray produces. But with the development of excellent resolution Compton camera systems this situation had changed. New concept of imaging using triple coincidence data was proposed~\cite{grignon2007nuclear}. In this new imaging, Compton cone drawn using extra gamma interaction points was used to estimate original annihilation point on the LOR, similar to TOF-PET imaging~\cite{giovagnoli2020pseudo}. Application of these type of radioisotopes is in generally of two types. It is used as conventional radiopharmaceutical, e.g., [44Sc]Sc-PSMA-617~\cite{eppard2018pre} in prostate cancer imaging. Besides, various non-pure beta emitters are formed as beam irradiation byproduct in ion therapy~\cite{parodi2007clinical}. Hence, it is online or offline beam range monitoring agent. But in this case, generally the emitted count is 1-3 orders magnitude smaller than conventional PET scan~\cite{kurz2015investigating}. On the other hand, it is known that tissue scattering can contribute to 30-60\% of coincidences in human 3D-PET~\cite{zaidi2007scatter} in which 80\% are single scattered~\cite{bailey2005positron}. So, in this work, we have investigated the feasibility of image reconstruction from those tissue scattered events in Compton-PET system. Our aim is to produce a physically meaningful image from the single scattered data which is independent from unscattered data. We believe having two independent image of same underlying activity distribution will assists us in better diagnosis. In this context, it is worth to mention that the motivation of WGI imaging concept is indeed to use all types data independently~\cite{yoshida2020whole}.\par

We performed GATE~\cite{jan2004gate} simulation with finite resolution parameters for a Compton-PET system with silicon as scatterer ring and LaBr$_3$:Ce as absorber ring. Geometrical arrangement and parameters were chosen keeping in mind the sensitivity and resolution. Line sources of $^{44}$Sc was used. And a cylindrical water phantom of diameter 10 cm was placed axially. At first, we had shown that the locus of tissue scattering point of a single scattered coincident (single scatter surface) is prolate spheroid (for scattering angle, ${\theta}_s < {90}^{0}$) and spindle toroid (for ${\theta}_s > {90}^{0}$) where to acquire these single scattered coincidences photo-peak and off-peak energy windows positioned in accordance with detector resolution were used. Data acquisition was performed using an appropriately defined trigger logic. Compton cones from both end of 511 keV detection were projected on the single scatter surface to obtain two 3D curves which cut each other in generally at two points forming two possible broken LORs. Finally, annihilation point was estimated by projecting Compton cone of 1157 keV and TOF information was used to choose between the two. The image we obtained is physically meaningful and proves the feasibility of single scattered imaging in non-pure beta emitter cases.

\section{Materials and Methods}

At first, we have discussed the locus of single scattering point in case of Compton-PET system. For proving the feasibility a GATE simulation was performed. Trigger logic was developed for data extraction. Finally the image reconstruction algorithm was proposed for single scatter imaging.

\subsection{Locus of scattering point}

We have drawn a typical Compton-PET set in figure (1). From here on, we named the locus as single scatter surface to avoid any confusion. Now, we assume a single scattered coincident event where annihilation happened at point O and tissue scattering at C. To find the locus of the scattering point C, at first, we write down the equation of locus depending only on scattering angle in tissue (${\theta}_s$),
\begin{equation*}
    \overrightarrow{AC}.\overrightarrow{CB}=\left | \overrightarrow{AC} \right |\left | \overrightarrow{CB} \right | \ cos \ {\theta}_s 
\end{equation*}
\begin{equation}
\label{eq.1}
    \Rightarrow \left ( \overrightarrow{r}-\overrightarrow{r_A} \right ).\left ( \overrightarrow{r_B}-\overrightarrow{r} \right )=\left | \overrightarrow{r}-\overrightarrow{r_A} \right |\left | \overrightarrow{r_B}-\overrightarrow{r} \right | \ cos \ {\theta}_s 
\end{equation}
where tissue scattering angle is calculated using this equation
\begin{equation*}
    {\theta}_s=\arccos \left ( 2-\frac{511}{{E_1}+{E_2}} \right )
\end{equation*}
where $E_1$ and $E_2$ are energies deposited in scatterer and absorber by the tissue scattered photon and here we have assumed a full energy deposition.\par

\begin{figure}[H]
    \centering
    \includegraphics[width=0.4\textwidth, height=0.4\textheight,keepaspectratio]{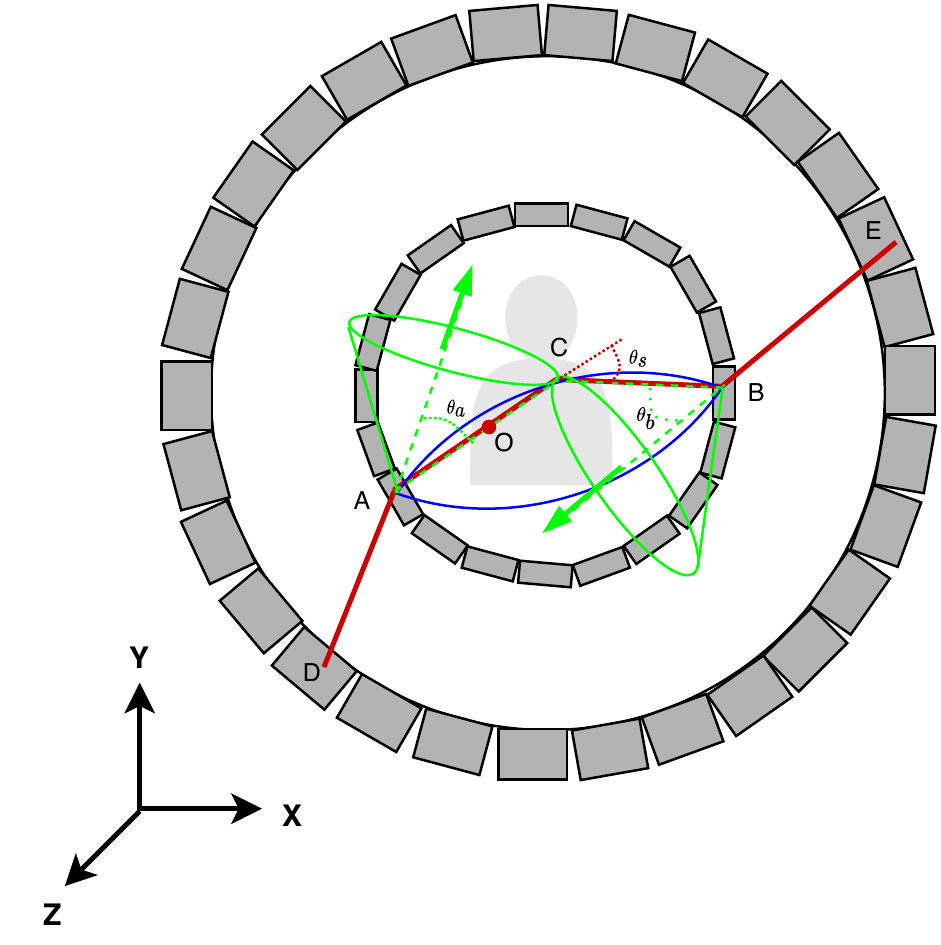}
    \caption{Compton-PET set up with a single scatter event get detected at points A, B (scatterer ring detection points) and points D, E (absorber ring detection points) where annihilation happended at O and tissue scattering happended at point C; tissue scattering angle is ${\theta}_s$ and scattering angles in scatterer ring are ${\theta}_a$, ${\theta}_b$; locus of scattering point shown in blue curve and Compton cones from both ends of 511 keV detection are shown in green colored cones.}
    \label{fig:GATE_simu}
\end{figure}

Now applying Compton cone constraint from both side of 511 keV detection the tissue scattering point can be further localised. The equations of the Compton cones are 

\begin{equation}
\label{eq.2}
    \left ( \overrightarrow{r}-\overrightarrow{r_A} \right ). \ \widehat{n_A}=\left | \overrightarrow{r}-\overrightarrow{r_A} \right | \ cos \  {\theta}_a
\end{equation}
and,
\begin{equation}
\label{eq.3}
    \left ( \overrightarrow{r}-\overrightarrow{r_B} \right ). \ \widehat{n_B}=\left | \overrightarrow{r}-\overrightarrow{r_B} \right | \ cos \ {\theta}_b 
\end{equation}
where ${\theta}_a$ and ${\theta}_b$ are scattering angles from scatterer ring and $\widehat{n_a}$ and $\widehat{n_b}$ are unit vectors along line joining from absorption to scattering point respectively.\par

It is known that eq. (1), which represents single scatter surface (blue curve in figure 1), is a surface equation of a prolate spheroid for ${\theta}_s < {90}^{0}$ and of spindle toroid for ${\theta}_s > {90}^{0}$. For further constraining the locus of scattering point, we have solved eq. (1) with eq. (2) and (3) which means solution between single scatter surface and cones. We found that the solution to be closed contour 3D curves on the single scatter surface. And two curves from both end of 511 keV detection cut each other in generally at two points. Further discussion about this is given in section 3.

\subsection{GATE simulation}

We performed a GATE~\cite{jan2004gate} simulation of a Compton-PET system. Silicon scatterer ring of thickness 2.5 cm and radius 20 cm was chosen. Radius was chosen larger since we are working with human scanner. And LaBr$_3$:Ce absorber of ring radius 28 cm and thickness 3 cm was used. Axial width of each ring was 28 cm. Energy resolutions of scatterer and absorber were 2.5\% and 5\% @511 keV and time resolutions were 1 ns and 200 ps respectively. Finally the spatical resolution was chosen to be 2 mm and 5 mm respectively. For image resolution study a $^{44}$Sc line source, situated at the centre of the scanner, of activity 1 MBq was used. Activity was chosen low to have a smaller number of random events. A cylindrical water phantom of diameter 10 cm and height 28 cm was defined axially. The decision of working with a human scale Compton-PET set up was due to the fact that the scatter fraction in human PET scan is significant enough to interest us in the proposed idea whereas in small animal imaging scatter fraction is not so high.

\subsection{Trigger logic}

After generating the data from GATE simulation, we had defined a trigger logic to select out usable valid triple gamma single scattered events. A coincidence time window of 10 ns was used for data selection. At first, two different energy windows, for 511 keV, the energy window was from 10-255 keV and for 1157 keV gamma it is 255-818 keV were used to select out scatterer detector interactions. If three hits in the above specified energy windows (two for 511 keV and one for 1157 keV) for the scatterer were obtained then we collected all the events in absorber ring falling in that coincidence time window and sorted out events with only three hits in absorber ring. In next stage, correspondence between individual scatter hit and absorber interactions were made. At first, the absorber hit corresponding to 1157 keV is identified depending on closeness of summed energy to 1157 keV. Then remaining two absorber hits were allocated depending on closeness from scatter hits. Finally, single scattered coincidences were acquired using photo-peak and off-peak energy windows of 495-525 keV and 250-495 keV respectively.

\subsection{Image reconstruction}

Image reconstruction was performed without applying any typical algorithm (e.g., MLEM, OSEM~\cite{bailey2005positron}). Rather the annihilation points were estimated independently for each event. At first, we had calculated two possible scattering points on the single scatter surface as explained in section 2.1. Then Compton cone of 1157 keV was projected on the two separate broken LORs to obtain at most four possible annihilation points (figure 2). One point among those was selected depending on FOV constraint and TOF information. 

\begin{figure}[H]
    \centering
    \includegraphics[width=0.4\textwidth, height=0.4\textheight,keepaspectratio]{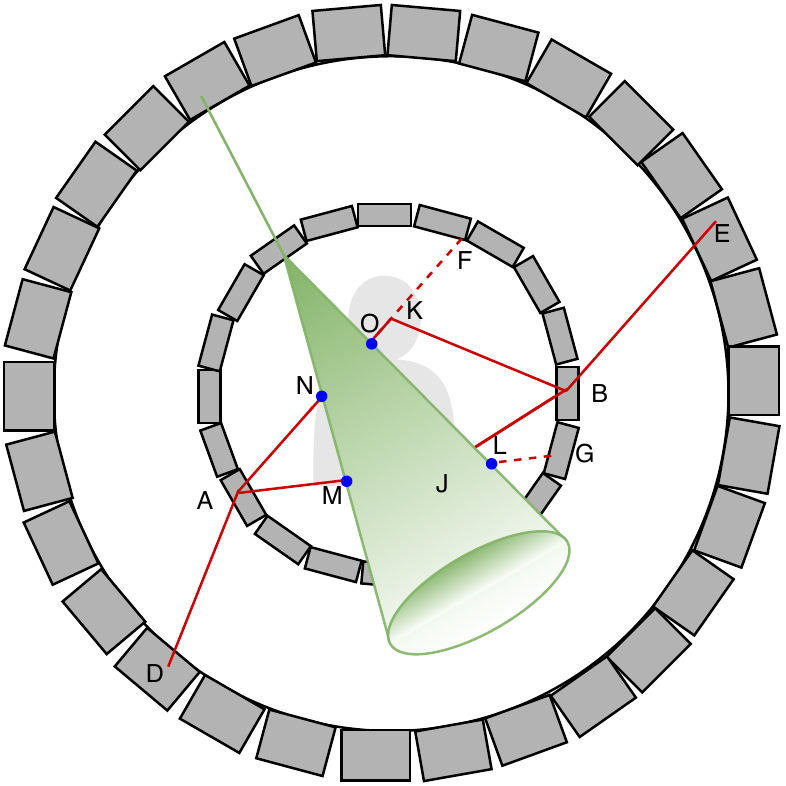}
    \caption{A single scattered event is detected at points A, B in scatterer ring and at point D, E in absorber ring; the intersection between single scatter surface, two Compton cones for 511 keV lefts us with two possible scattering points K, J; the intersection between Compton cone of 1157 keV and broken LORs gives us four possible annihilation points; one among those are chosen depending on FOV constraint and TOF information.}
\end{figure}

\section{Results}

As described in the section 2.1, using Compton cones from both sides of 511 keV detection, the locus can be further constrained. The cross section between single scatter surface and the cone in these cases is a type of 3D curves (red and green curve in figure 3) such that two such 3D curves cut each other in generally at two points. It is worth to mention here that the generation of two cross points is not due to finite resolution of detectors and hence those can be quite a distant apart (figure 3). 

\begin{figure}[H]
    \centering
    \includegraphics[width=0.6\textwidth, height=0.6\textheight,keepaspectratio]{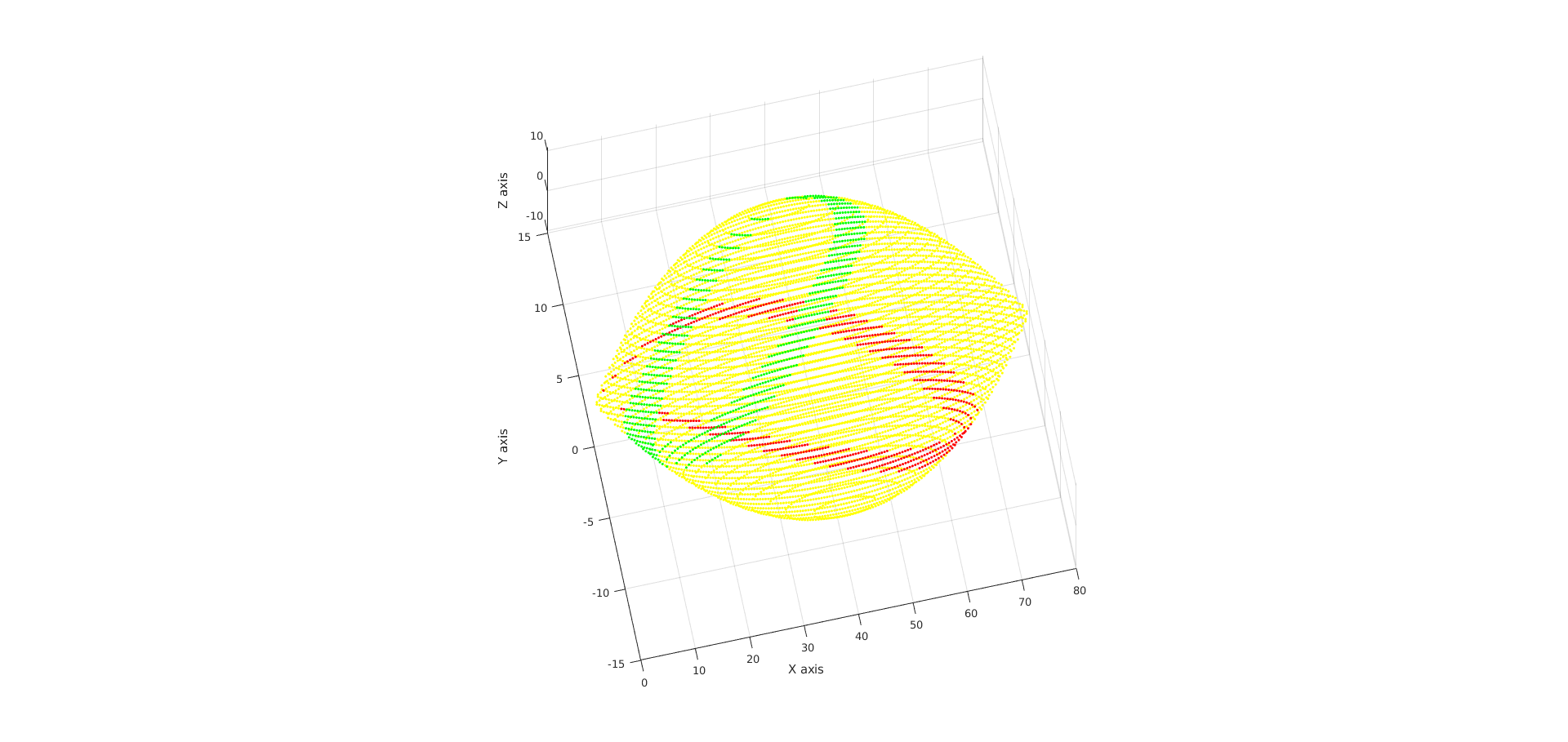}
    \caption{The solution between single scatter surface (yellow envelope) and Compton cones from each end of 511 keV detection is shown as green and red curve, in generally these two curves cut each other at two points, it is to be noted that two point is not due to finite resolution of detector.}
\end{figure}

We had performed the GATE simulation of Compton-PET system with parameters described in section 2.2. Then root output data was processed using the trigger logic described in section 2.3. The trigger logic was implemented through MATLAB scripts. Figure 4-6 shows the 2D energy histogram plot between scatter energy deposition vs. absorber energy deposition for 511 and 1157 keVs. For unscattered photons we can find x+y=511 keV line where x and y are absorber and scatter deposition respectively which shows that the proposed trigger logic is able to collect 511 keV data (figure 4). The line is discontinued at scatter energy 10 keV and 255 keV, because of energy window on scatter deposition (see section 2.3). Besides the width of the x+y=511 line is decided by photo-peak width chosen in trigger logic. On the other hand, for 1157 keV detection similar x+y=1157 keV line can be seen (figure 5). Here we have discontinuity on scatterer energy at 255 keV and 818 keV due to energy window applied in trigger logic. Point to be noted, here unlike 511 keV we have count below the x+y=1157 line. This is because there is no window applied on total energy like photo-peak energy window for 511 keV. We have assumed a full deposition of energy of 1157 keV gamma in scatterer and absorber. Finally, for single scattered events, rather than having a line we have area bounded by x+y=250 keV, x+y=495 keV, x=10 keV, and x=255 keV (figure 6). First two bounds are due to off-peak window and last two are applied in initial stage of trigger logic. 

\begin{figure}[H]
\centering
\includegraphics[width=0.5\textwidth, height=0.5\textheight,keepaspectratio]{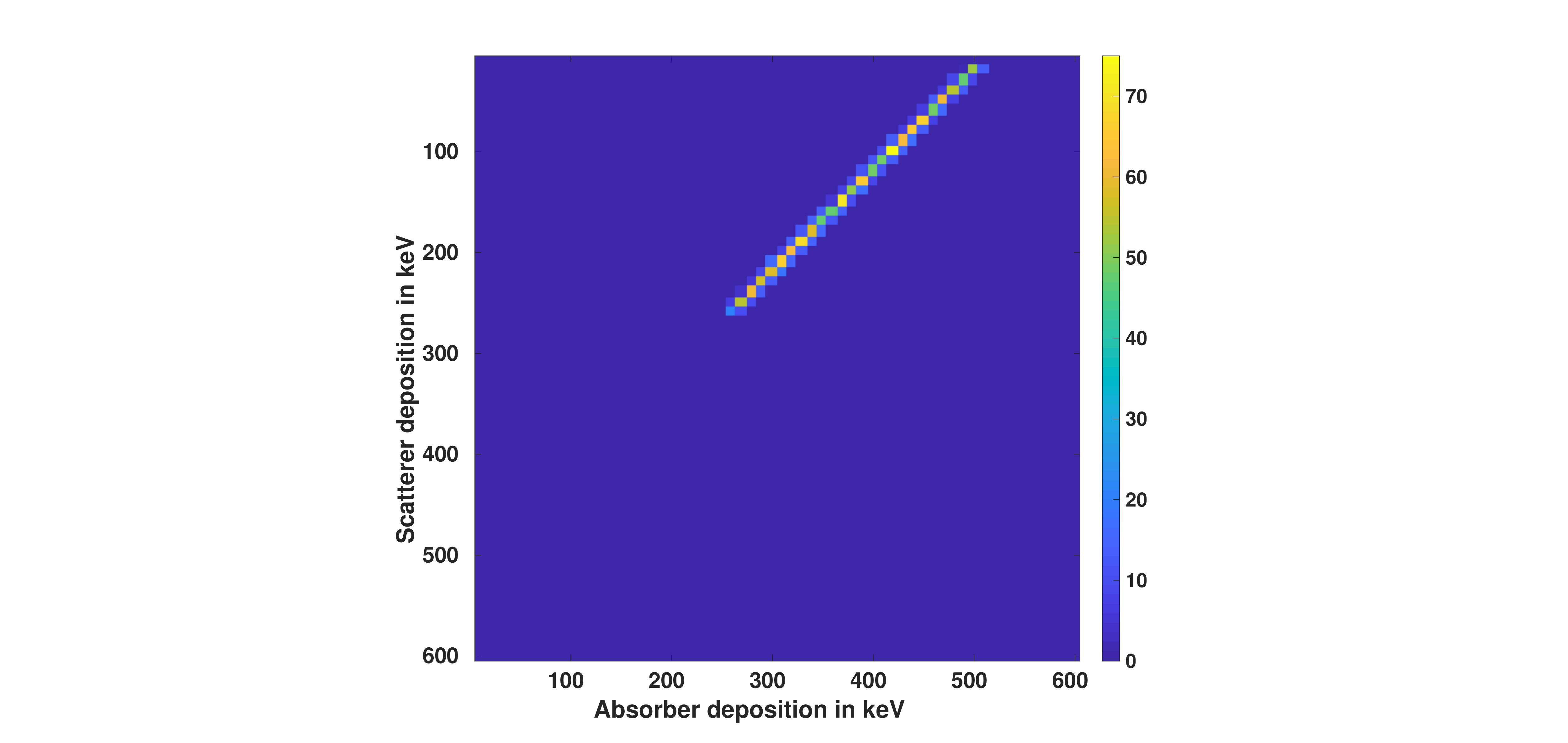}
\caption{The count histogram color plot between scatterer and absorber energy deposition for 511 keV unscattered photon detection.}
\end{figure}

\begin{figure}[H]
\centering
\includegraphics[width=0.5\textwidth, height=0.5\textheight,keepaspectratio]{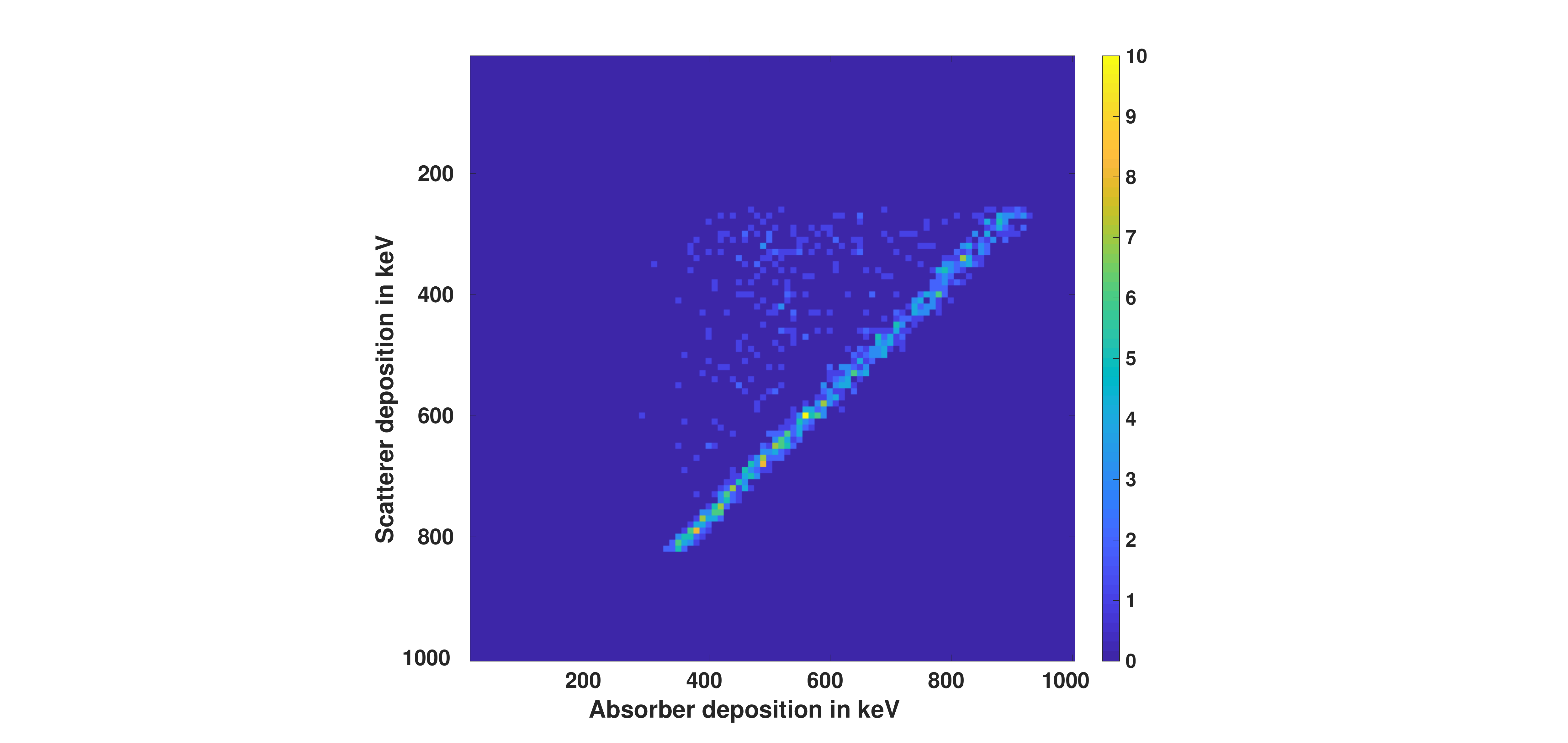}
\caption{The count histogram color plot between scatterer and absorber energy deposition for 1157 keV photon detection.}
\end{figure}

\begin{figure}[H]
\centering
\includegraphics[width=0.5\textwidth, height=0.5\textheight,keepaspectratio]{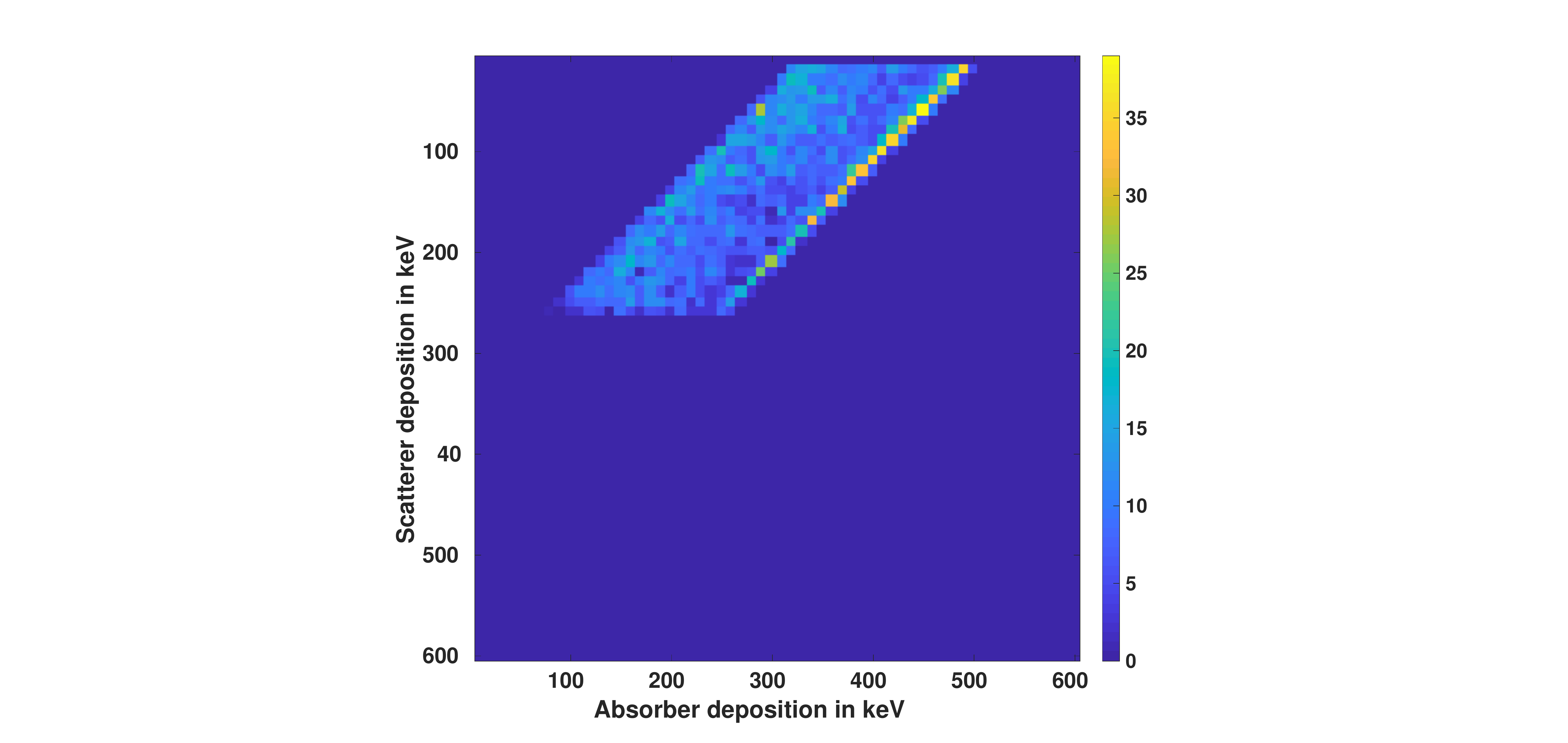}
\caption{The count histogram color plot between scatterer and absorber energy deposition for 511 keV single scattered photon detection.}
\end{figure}

\begin{figure}[H]
\centering
\includegraphics[width=0.5\textwidth, height=0.5\textheight,keepaspectratio]{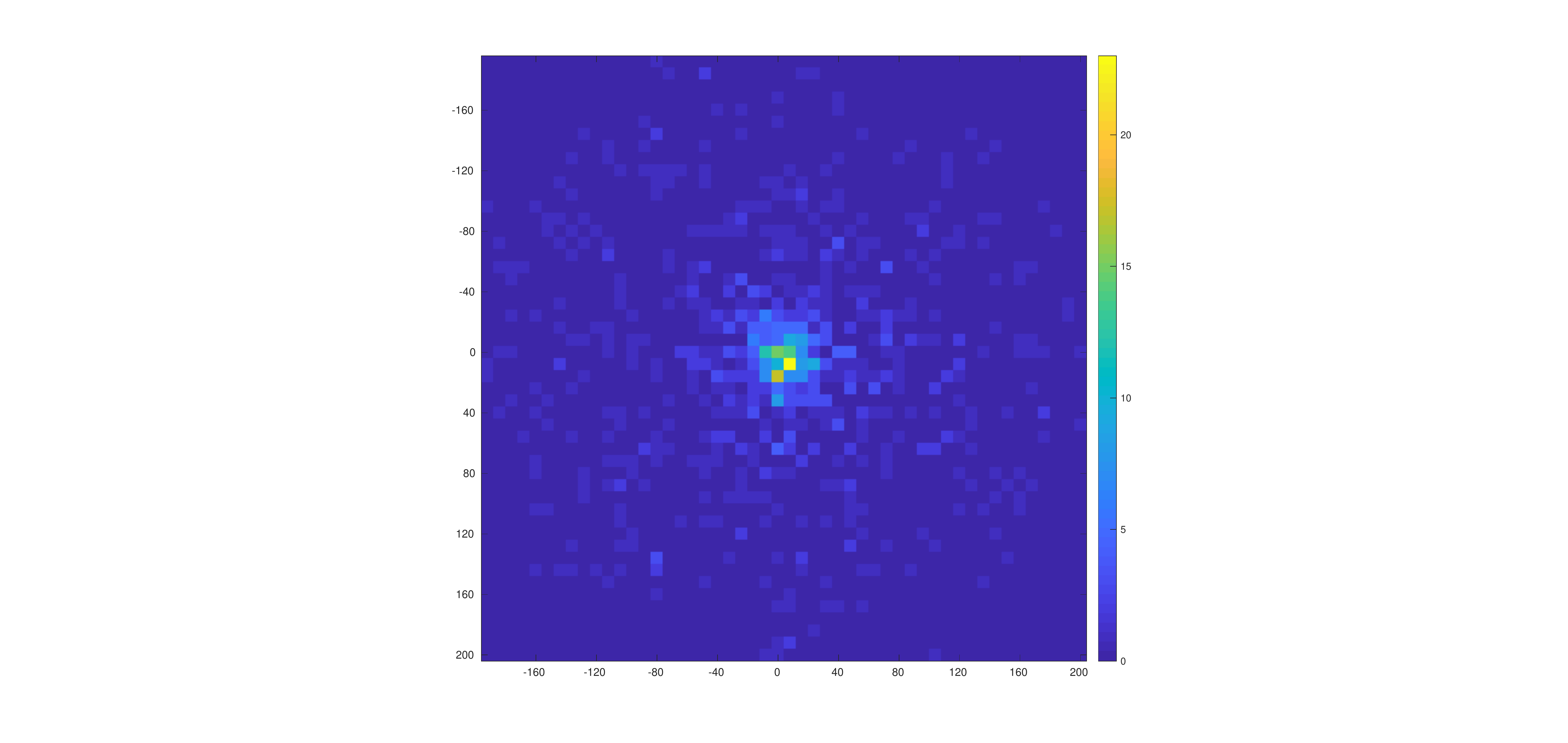}
\caption{Single scattered image (cross sectional) of line source, pixel size was chosen to be $8\times 8$ $mm^2$. }
\end{figure}

\begin{figure}[H]
\centering
\includegraphics[width=0.3\textwidth, height=0.3\textheight,keepaspectratio]{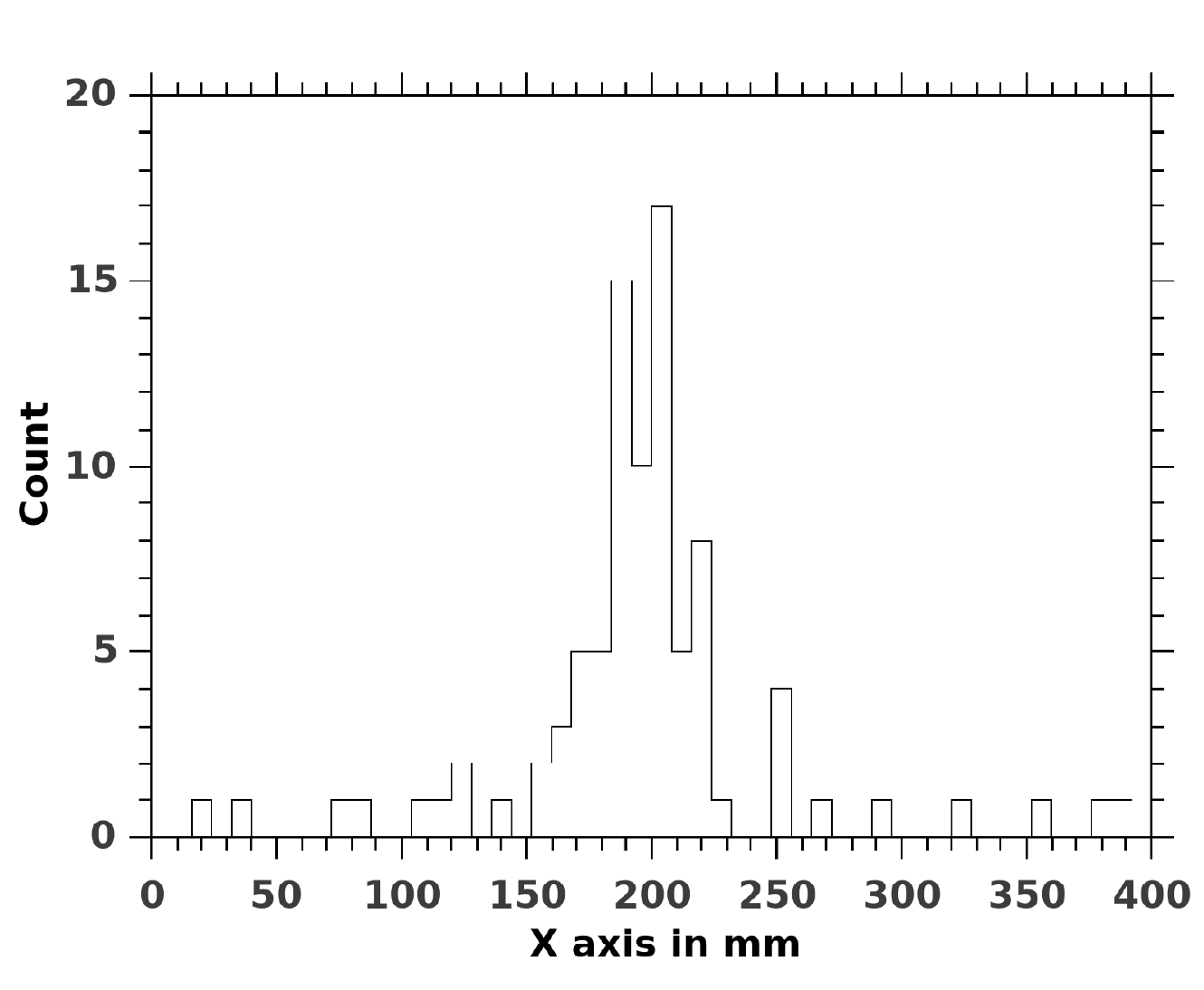}
\caption{Intensity line profile (horizontal) of the single scatter image through (0,0) point, pixel size was chosen to be $8\times 8\ mm^2$.}
\end{figure}

Finally, we had produced image from single scattered data set (figure 7). For image resolution study, we had calculated FWHM of intensity line profile (horizontal) of the image. The histograms were shown in figure (8) with FWHM calculated to be 35.864 mm. To sum up, we were able to produce physically meaningful single scatter images. This proves the feasibility of single scatter imaging for Compton-PET system with triple gamma source.

\section{Conclusion}

We have proposed the idea of feasibility of imaging from single scattered (inside tissue) data in triple gamma imaging. Tissue scattered data in human PET scan can go up to 40-60\%. On the other hand, triple gamma imaging suffers from low count specially in online ion range verification in ion therapy and hence in that context the idea of imaging from scattered data is relevant. Although a better resolution image than unscattered image can’t be expected from scattered data due to inherent resolution effects, we believe that producing image from two independent data sets -- unscattered and single scattered -- will improve our diagnosis ability. We have shown the feasibility of the proposed concept. Analysing GATE simulation data, we are able to produce physically meaningful images. The trigger logic used here is not claimed to be perfect. Rather simplicity is invoked to make the task computationally simple as this work is related to only feasibility. We believe that the idea proposed can be beneficial in triple gamma imaging based beam range monitoring and late point imaging in case of Scandium DOTA-TOC imaging.

\section{Acknowledgement}

Authors wish to gratefully acknowledge the Center for Development of Advanced Computing (C-DAC), Pune, India, for providing the supercomputing facilities~\cite{yuvaindia} for data analysis.

\printbibliography

\end{document}